\documentclass[useAMS,usenatbib]{mnras}

\usepackage{url,times,graphicx,hyperref,amsmath,amsfonts,amssymb,aas_macros,color,epsfig,epstopdf}

\newcommand{\vtwo}{$V_{\rm 2kpc}$}
\newcommand{\vrlast}{$V_{\rm Rlast}$}

\newcommand{\lcdm}{$\Lambda$CDM}
\newcommand{\msun}{M$_\odot$}

\newcommand{\kms}{km$\,$s$^{-1}$}

\newcommand{\hMpc}{{\ifmmode{h^{-1}{\rm Mpc}}\else{$h^{-1}$Mpc}\fi}}
\newcommand{\hkpc}{{\ifmmode{h^{-1}{\rm kpc}}\else{$h^{-1}$kpc}\fi}}
\newcommand{\hMsun}{{\ifmmode{h^{-1}{\rm {M_{\odot}}}}\else{$h^{-1}{\rm{M_{\odot}}}$}\fi}}
\newcommand{\ltsima}{$\; \buildrel < \over \sim \;$}
\newcommand{\gtsima}{$\; \buildrel > \over \sim \;$}
\newcommand{\lsim}{\lower.5ex\hbox{\ltsima}}
\newcommand{\gsim}{\lower.5ex\hbox{\gtsima}}
\def\LCDM{$\Lambda$CDM}

\def\lesssim{\mathrel{\hbox{\rlap{\hbox{\lower4pt\hbox{$\sim$}}}\hbox{$<$}}}}
\def\gtrsim{\mathrel{\hbox{\rlap{\hbox{\lower4pt\hbox{$\sim$}}}\hbox{$>$}}}}

\newcommand{\Sec}[1]{Section~\ref{#1}}

\newcommand{\beq}{\begin{equation}}
\newcommand{\eeq}{\end{equation}}
\def\beqa{\begin{eqnarray}}
\def\eeqa{\end{eqnarray}}
\def\hMpc{$h^{-1}\,{\rm Mpc}$}
\def\hkpc{$h^{-1}\,{\rm kpc}$}
\def\LCDM{\ensuremath{\Lambda}CDM}

\def\head{
 \vbox to 0pt{\vss
                   \hbox to 0pt{\hskip 440pt\rm LA-UR-10-07069\hss}
                  \vskip 25pt}}

\title[Reproducing the diversity of dwarf galaxy RCs]
{NIHAO XIV: Reproducing the observed diversity of dwarf galaxy rotation curve shapes in \lcdm}
\author[I. Santos-Santos et al.]
       {Isabel M. Santos-Santos$^{1,2}$\thanks{E-mail: isabelm.santos@uam.es}, Arianna Di Cintio$^{3}$\thanks{Karl-Schwarzschild fellow}, Chris B. Brook$^{4,5}$\thanks{Ram\'on y Cajal Fellow}, Andrea Macci\`o$^{6,7}$,
        \newauthor  Aaron Dutton$^{6}$ \& Rosa Dom\'inguez-Tenreiro$^{1,2,8}$ 
          \\
$^{1}$Departamento de F\'isica Te\'orica, Universidad Aut\'onoma de Madrid, E-28049 Cantoblanco, Madrid, Spain\\
$^{2}$Astro-UAM, UAM, Unidad Asociada CSIC, E-28049 Cantoblanco, Madrid, Spain\\
$^{3}$Leibniz Institute for Astrophysics Potsdam (AIP), An der Sternwarte 16, D-14482 Potsdam, Germany\\
$^{4}$Departamento de Astrof\'isica, Universidad de La Laguna, Av. del Astrof\'isico Francisco S\'anchez s/n, E-38206 La Laguna, Tenerife, Spain\\
$^{5}$Instituto de Astrof\'isica de Canarias, C/ V\'ia L\'actea s/n, E-38205 La Laguna, Tenerife, Spain\\
$^6$New York University Abu Dhabi, PO Box 129188, Saadiyat Island, Abu Dhabi, UAE\\
$^7$Max-Planck-Institut f\"ur Astronomie, K\"onigstuhl 17, D-69117 Heidelberg, Germany\\
$^8$Centro de Investigaci\'on Avanzada en F\'isica Fundamental, Universidad Aut\'onoma de Madrid, E-28049 Cantoblanco, Madrid, Spain
}

\begin{document}

\date{Accepted XXXX . Received XXXX; in original form XXXX}

\pagerange{\pageref{firstpage}--\pageref{lastpage}} \pubyear{0000}

\maketitle

\label{firstpage}


\begin{abstract}
The significant diversity of rotation curve (RC) shapes in dwarf galaxies  has recently emerged as a  challenge to  \lcdm: in dark matter (DM) only simulations,  DM halos have a universal \textit{cuspy} density profile that results in self-similar RC shapes. 
We  compare RC shapes of simulated galaxies from the NIHAO project with observed galaxies from the homogeneous SPARC dataset.
The DM halos of the NIHAO galaxies can expand to form \textit{cores}, with the degree of expansion depending on their stellar-to-halo mass ratio. 
By means of the $V_{\rm 2kpc}-V_{\rm Rlast}$ relation (where \vrlast\ is the outermost measured rotation velocity),
 we show that both the average trend and the scatter in RC shapes of NIHAO galaxies are in reasonable agreement with SPARC: this represents a significant improvement compared to simulations that do not result in DM core formation, 
suggesting
 that halo expansion is a key process in matching the  diversity of  dwarf galaxy RCs. 
Note that NIHAO galaxies can reproduce even the extremely slowly rising RCs of IC~2574 and UGC~5750.
Revealingly, the range where observed galaxies show the highest diversity corresponds to the range where core formation is most efficient in NIHAO simulations, 50$<$\vrlast/\kms$<$100.
A few observed galaxies in this range cannot be  matched by any NIHAO RC
nor by simulations that predict a universal halo profile.
 Interestingly, the majority of  these are starbursts or emission-line galaxies, with steep RCs and  small effective radii.
  Such galaxies represent an interesting observational target  providing new clues to
    the process/viability of \textit{cusp-core} transformation, 
    the  relationship between starburst and inner potential well, 
  and the nature of DM.

\end{abstract}

\noindent
\begin{keywords}
 galaxies: evolution -- formation -- haloes cosmology: theory -- dark matter
 \end{keywords}

\section{Introduction} \label{sec:introduction}


A well established outcome of the $\Lambda$-cold dark matter (\LCDM) cosmological model is the self-similarity of dark matter (DM) halos that host galaxies: their density distribution is described by a steep, `cuspy' profile, introduced 20 years ago by \cite{navarro97}.
Although numerical simulations based on the \LCDM\ paradigm can reproduce the clustering of cosmic structures on large scales \citep{Klypin16}
and the two and three-point correlation functions of galaxies \citep{Guo16,Guo16b,RodriguezTorres16}, these cuspy halos are at odds with the internal structure of many dwarf galaxies, as inferred by observations of their rotation curves (RCs).  This is a decades-long  problem  known as the `cusp-core' discrepancy \citep{Moore94, Flores94, DeBlok08,DeBlok10,Oh15}.
In fact, observations of  rotation curves (RCs) of low-mass galaxies reveal a great diversity in their shapes \citep{Oman15}. Many dwarf RCs rise slowly and gently toward the galaxy's outskirts, indicative of `cored' DM density profiles,  while others rise rapidly, more in line with  cuspy profiles. 

The low baryonic mass fraction of dwarf galaxies makes it tempting to assume that their DM halos can be directly related to results from DM-only simulations. However,   hydrodynamical simulations indicate that the DM halos of dwarf galaxies can be significantly modified by baryonic processes \citep[e.g.,][]{governato10,Governato12,DiCintio14a,Cloet14,chan15,Verbeke15,Tollet16,Read16}.
Some of these hydrodynamical simulations  include feedback processes 
such as energy released by young and massive stars  \citep{stinson13} and winds projected from supernovae explosions \citep{stinson06},  which result in outflows of gas from the inner region of galaxies.
The simulations show how these outflows can disperse DM particles, reshaping the  underlying DM distribution as the galaxy evolves  \citep{Navarro96b}, and induce core creation when repeated outflow events occur as a result of a bursty star formation \citep{Read05, pontzen12}.
This hypothesis gains support by 
observational evidence for strong gas outflows from the centres  of galaxies at all redshifts \citep{Shapley03,Martin05, Vanderwel11},
as well as by  studies on the star formation histories of local and  nearby dwarf galaxies \citep[e.g.,][]{McQuinn10a, Lelli14c, Weisz14,Izotov16}.
In addition, strong outflows have been integral in the progress toward simulating realistic populations of galaxies \citep[][]{brook12,munshi13,hopkins14,Wang15,schaye15,Agertz15}.

The issue  of the large variety of dwarf RC shapes has been explored recently  using different  models.  
In particular, {\citet{Oman15} studied the RC shapes  of simulated galaxies versus a compilation of observed galaxies by comparing   the circular velocities at the maximum of the RC, $V_{\rm max}$, and in the inner region of the galaxy disc, \vtwo, and they noted the wide range of shapes --or \vtwo\ values for a given $V_{\rm max}$-- arising in observations. 

\cite{brook15a} showed with a semi-empirical model that 
by
combining the  scatter found in observed galaxy scaling relations with the scatter in the halo mass-concentration relation, 
one
can
obtain
 the required  scatter in RC shapes; assuming that DM halos  follow the DC14 mass dependent density profile \citep{DiCintio14b} and hence are cored at a particular mass range, $M_{\rm star}\sim10^{6.5-9.5}$ \msun.
This result reflects the fact that baryons contribute significantly to the inner mass of cored profiles,  so 
\textit{the observed diversity in disc scale-lengths will result in a higher  diversity of RC shapes} 
than the case of a cuspy profile, in which the contribution of baryons to the central mass is  lower.
The result still begs the question of the origin of the diversity of scale lengths and the relation between baryonic and DM mass distributions. 

\citet{Read16} used 
very high-resolution
hydrodynamical simulations of isolated dwarf galaxies to show that different RC shapes
and DM density profiles
 may be  a reflection of the star formation stage the galaxy is under-going: either bursty, post-bursty, or quiescent; that pushes the amplitude of the RC from steeply rising, to very shallow rising, to something in between, respectively. 
These simulations indicated that the steep RCs of starburst galaxies are caused by the increased gas turbulence driven by the star formation.

\citet{Katz17} explored the issue in a complementary manner,  building  mass models for observed galaxies with reliable stellar mass  measurements,  
assuming a cuspy NFW \citep{navarro97} and a modified-by-baryonic-feedback  DC14 \citep{DiCintio14b} 
halo profile 
to fit  their observed RCs.
The results showed that
 using an expanded halo profile with the degree of expansion depending on the stellar-to-halo mass ratio \citep{DiCintio14a}
allowed  the recovery of the observed RCs for a very broad range of galaxy masses and surface brightnesses, and, additionally, allowed to match  the $M_{\rm star}/M_{\rm halo}$ and mass-concentration relations
as predicted by abundance-matching and \lcdm\ theory.

Dark matter alternative models like those of self-interacting dark matter  (SIDM) have previously shown to produce cored DM halos \citep{Rocha13}.
More recently, controlled N-body simulations that used SIDM particles and mimicked the effect of a baryonic disc were also able to better reproduce the observed  distribution of dwarf galaxy RC shapes than CDM-only simulations  \citep{Creasey17}, establishing that the combination of SIDM and baryonic effects may represent a viable solution to the problem. This is in agreement with previous analytical calculations within the SIDM scenario made by \cite{Kamada16} and with hydro-SIDM simulations from \citet{fry15}.
However, in order to increase the scatter in RC shapes and match mass profiles from dwarf galaxies to cluster scales, SIDM models require velocity dependent cross-sections, invoking 
additional parameters beyond \lcdm.
A more robust examination of the validity of this model would be possible from a self-consistent simulation of SIDM with baryons and its prediction for different halo mass scales.



In this paper we  compare the RC shapes of  the new SPARC  dataset \citep{Lelli16}, a collection of high-quality 
H\,{\sc i}/H~$\alpha$ RC data of nearby galaxies that span a wide range in luminosities,  to state-of-the-art  hydrodynamical  zoom-in simulations from the NIHAO project \citep{Wang15},  that span a wide range in masses and merger histories. 
NIHAO uses a cold DM (CDM) model, and  includes a  feedback implementation that naturally produces 
galaxies hosted in halos that have undergone
a wide spectrum of halo responses, from contraction to expansion. 
 The degree of expansion
has been shown to 
  depend
primarily 
 on the stellar-to-halo mass ratio \citep{DiCintio14a,Tollet16},
 and then eventually, at a given stellar mass, also on size \citep{Dutton16}, with larger galaxies being more susceptible to expansion.
  Following the method used in \citet{Oman15}, we  compare the circular velocities at the inner and outermost parts of the RCs to define their shapes and then assess their diversity, paying particular attention to  mass dependence.

The paper is organized as follows. The suites of observed and simulated galaxies are introduced in Secs. \ref{sec:sims} and \ref{sec:obs}.
In Sec. \ref{comparing} we present the method used for
obtaining an equivalent radius at which to measure the circular velocity in galaxies of similar mass.
The comparison of RC shapes and its discussion is done in Sects. \ref{matching} and \ref{notmatched}, and finally the results are summarized in Sec. \ref{sec:conc}.


\section{Methods}\label{sec:methods}

\subsection{Simulations}\label{sec:sims}
We use the NIHAO (Numerical Investigation of a Hundred Astrophysical Objects) project \citep{Wang15}  suite of $\sim$100 hydrodynamical zoom-in simulations of isolated halos, with halo masses ranging from $M_{\rm halo}= 4\times10^9 - 3 \times 10^{12}$ M$_\odot$, and run in a flat \lcdm\ cosmology with parameters from \citet{Planck14}:
$H_0= 67.1$ km s$^{-1}$
Mpc$^{-1}$, $\Omega_{\rm m} = 0.3175$,
$\Omega_\Lambda= 1 - \Omega_{\rm m} - \Omega_{\rm r} = 0.6824$,
 $\Omega_{\rm r} = 0.00008$,
 $\Omega_{\rm b} = 0.049$,
$\sigma_8 = 0.8344$, $n = 0.9624$,
  and initial conditions created
using a modified version of the grafic2 package as described in \citet{Penzo14}.

The refinement level in the different simulation boxes is chosen in order to maintain a roughly
constant relative resolution (i.e., $\epsilon_{\rm DM}/R_{\rm vir} \sim  0.003$), with
$\sim 10^6$ DM particles per halo. This allows the resolution of the mass profile to below 1 per cent of the virial radius at all halo masses 
ensuring that galaxy half-light radii are well resolved. For the specific values of resolution see \citet{Wang15}.

The hydrodynamical run has been done with a version of the N-body SPH code GASOLINE \citep{Wadsley04,Keller14}, referred to as 
ESF-GASOLINE2 \citep{Wang15}.
As in the original version of GASOLINE, it has prescriptions for gas hydrodynamics and cooling,
hydrogen, helium and metal-line cooling \citep{Shen10}, photo ionization and heating from a \citet{Haardt96} UV background, star formation according to a $\rho^{1.5}$ Kennicutt-Schmidt Law where stars form when gas particles are cold ($T<15000$ K)
and dense ($n_{\rm th}>10.3$ cm$^{-3}$),
 supernova feedback following the blastwave formalism \citep{stinson06}, `early stellar feedback' from massive stars \citep{stinson13}, 
 and metal diffusion. The new version includes an updated hydrodynamics solver, described in \citet{Keller14}.


The NIHAO simulations have already proved to reproduce realistic
galaxies matching many properties and scaling relations which we group
into three classes.  Baryonic content: the stellar mass vs halo mass
relation from halo abundance-matching and the star formation rate vs
stellar mass relation \citep{Wang15}; neutral gas vs stellar mass
relation \citep{Stinson15}; circum-galactic medium \citep{Gutcke17}; Tully-Fisher scaling relations \citep{Dutton17}; the baryon budget of the Milky Way \citep{Wang17}.  Dark matter
properties: the shape of the Milky Way's DM halo \citep{Butsky16}; central DM density slopes of dwarf galaxies \citep{Tollet16}; resolves the Too-Big-To-Fail problem for field
dwarf galaxies \citep{Dutton16a}; the HI velocity function
\citep{Maccio16}.  Galaxy structure: the kinematics of galaxy
disks \citep{Obreja16}; ultra diffuse galaxies \citep{DiCintio17}; clumpy morphology of high redshift galaxies \citep{Buck17}.

 For this work we have used those galaxies with total (inside 10 per cent of the virial radius) stellar masses in the $M_{\rm star} = 6\times 10^6-10^{11}$ M$_\odot$ range: a
 sample  of 73 simulated galaxies.

\begin{figure}
\hspace{0.cm}\includegraphics[width=\linewidth]{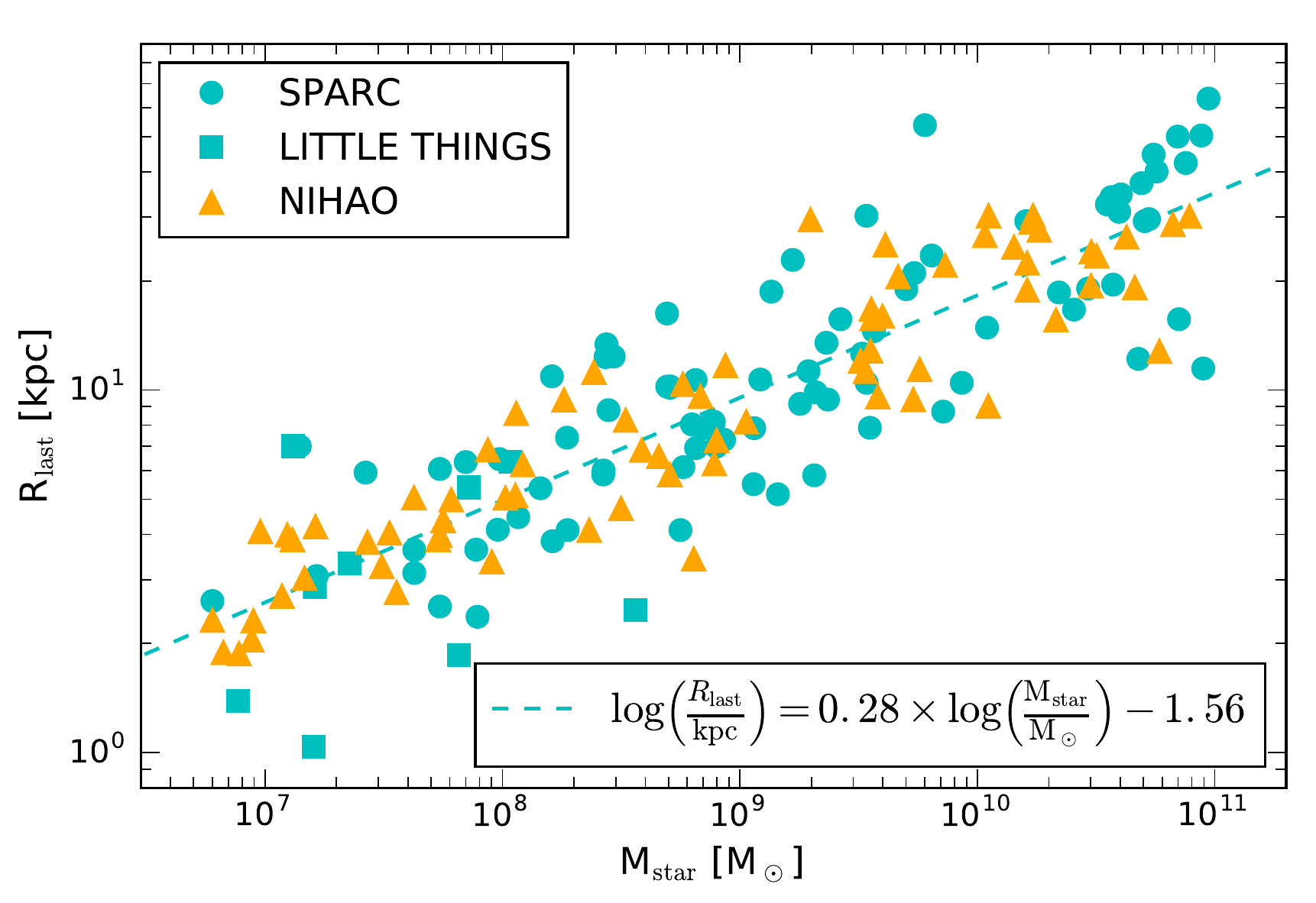}
\caption{The $M_{\rm star}-R_{\rm last}$ relation for observations and simulations. Data representing the SPARC galaxies are shown as cyan circles and are fitted by: 
$\log \left(\frac{R_{\rm last}}{\rm kpc}\right)=0.28 \log \left(\frac{M_{\rm star}}{\rm M_\odot}\right)-1.56$
 (dashed line), with a standard deviation of 0.17 dex. This linear relation and its scatter have been used to compute the corresponding $R_{\rm last}$ values for the NIHAO sample of galaxies according to their stellar masses. Simulated results are overplotted as orange triangles.
}
\label{fig:mstarrlast}
\end{figure}

\subsection{Observations}\label{sec:obs}

Observational galaxies are taken from the new SPARC (Spitzer Photometry \& Accurate RCs) dataset \citep{Lelli16}
and the LITTLE THINGS (Local Irregulars That Trace Luminosity Extremes, The HI Nearby Galaxy Survey) galaxies analyzed in \citet{Iorio17}.

SPARC is a sample of 175 nearby galaxies with homogeneous photometry at 3.6 $\mu$m and high-quality RC data from previous H\,{\sc i}/H~$\alpha$ studies.
Though not a complete sample,
it is representative of nearby disc galaxies 
spanning a very wide range in
 luminosities (therefore stellar masses), morphologies and surface brightnesses.

The LITTLE THINGS data used in this work is the one studied in \citet{Iorio17}, 
in which a sub-sample of 17 dwarf irregulars, taken from the original sample of \citet{Oh15}, was reanalyzed. 
These galaxies conserve the same statistical distribution of galactic properties as the original sample.
Rotation curves were built following the 3D-approach of the $^{\rm 3D}$BAROLO  software \citep{DiTeodoro15}, which allows results to be unaffected by beam smearing.

SPARC RCs 
have been corrected for  beam smearing, inclination, and pressure support (or asymmetric drift).
The \citealt{Iorio17}'s LITTLE THINGS RCs have also been carefully corrected for pressure support.
We note however that this correction is important in very few cases of low mass galaxies that have a velocity dispersion comparable to their rotational velocity.
Therefore, even though the corrections have been done differently for each sample \citep[see secs. 5.2 and 4.3 from][respectively]{Lelli14b, Iorio17}, we assume that the observed RC data used here effectively traces the circular velocity of the galaxy. 
We acknowledge that apart from pressure support, 
other deviations from equilibrium can be produced by the effect of factors such as HI bubbles from star formation events \citep{Read16a} or gas non-circular motions, which seem to be common near the centre of very low mass galaxies. 
Due to the complexity of the problem, as recently discussed in \citet{Oman17}, a thorough  study of these effects would be beyond the scope of the present paper.

For this study we select observed galaxies inside the same stellar mass range
 as simulated galaxies.
 For SPARC the stellar masses are computed assuming a constant stellar mass-to-light ratio of $M_{\star}/L_{[3.6]}=\Upsilon_{\star}=0.5$, value which has been shown to 
minimize the scatter around the
baryonic Tully-Fisher relation
 \citep[BTFR, ][]{Lelli16a}
 and to give a realistic galaxy population on the $f_{\rm gas}-\log(L_{3.6})$ relation
 \citep[see][]{Lelli16}.
For LITTLE THINGS we use the highest stellar mass between the two values given in \citet{Oh15}, computed from kinematics and SED fitting respectively, in order to include the maximum number of objects possible.
 Moreover,
 we use only those objects
  with inclinations greater than 45$^{\circ}$ to ensure the most reliable RC data\footnote{Note that including all
  galaxies with inclinations $\geq$30$^{\circ}$ does not substantially change our results.}.
  This leaves a sample of 
  94 observed galaxies,
  of which 85 belong to SPARC and 9 to LITTLE THINGS.


\subsection{Comparing observed and simulated rotational velocities} \label{comparing}

Various definitions for the `characteristic' rotational velocity of observed galaxies $V_{\rm rot}$ are often used: $V_{\rm flat}$, $V_{\rm max}$, $V_{\rm DHI}$, $W_{\rm 20}$, $W_{\rm 50}$... 
\citep[e.g.,][]{Verheijen01, Bradford15, Bradford16, McGaugh15},
 which can lead to confusion when comparing rotational velocities  of simulations with observations. By studying the different BTFRs obtained with various
 definitions,  we have formerly  
  highlighted the need to be consistent about the particular method of measuring $V_{\rm rot}$, and 
precise about the radius at which velocities are measured
\cite*[see fig. 5,][]{Brook16}.

Indeed, the commonly used $V_{\rm max}$ and $V_{\rm flat}$ measures are ambiguous, since many observed low mass galaxies have RCs that are still rising at their last measured point \citep[e.g.][]{kuziodenaray06,catinella06}, and the data does not reach the true maximum or flat part of their RCs
\citep{brookdicintio15, Sales17}.
Therefore, taking into account that observational RC data extends out to a certain, limited radius for each galaxy, while simulations allow calculating the RC out to any radius,
 we have searched in this work for an equivalent radius at which to measure $V_{\rm rot}$ in both observed and simulated galaxies with similar stellar masses.

We use the relation between $M_{\rm star}$ and $R_{\rm last}$ from the observational sample, 
where $R_{\rm last}$ is the radius at the last measured point of the RC.
  We choose to compare $R_{\rm last}$ with stellar mass because for all SPARC galaxies this quantity is estimated in a uniform manner from precise 
 near-infrared surface photometry ($L_{\rm 3.6 \mu m}$).
 In the case of simulations, the stellar mass is  unambiguously and directly obtained. 
 Using the total baryonic mass $M_{\rm bar}$ gives a slightly tighter relation, but
  the different ways of measuring the `cold' or H\,{\sc i} gas in observed and simulated galaxies introduces an extra uncertainty
 \citep[e.g.][]{Stinson15, Sales17}
 which we want to avoid.

We used a linear fit for the observed $M_{\rm star}-R_{\rm last}$ relation, with its scatter,
 to derive the corresponding $R_{\rm last}$ for a simulated galaxy of given stellar mass.
In Fig. \ref{fig:mstarrlast} we show as cyan points the $M_{\rm star}-R_{\rm last}$ relation for
the set of observed
 galaxies. A dashed line represents the linear fit to these points, which is explicitly shown in a box in the lower part of the figure. 
The last measured point of the RC increases with stellar mass as expected, 
following 
\begin{equation}
\log\left(\frac{R_{\rm last}}{\rm kpc}\right)=0.28 \times \log\left(\frac{M_{\rm star}}{\rm M_\odot}\right)-1.56
\label{eqrlastmstar}
\end{equation}
and presents a moderate scatter for a given stellar mass, which translates into a standard deviation in 
$\log(R_{\rm last})$
 of $\sigma=$ 
0.17 dex.
Applying this relation, we have computed the  corresponding $R_{\rm last}$ radii for the NIHAO galaxies according to their stellar mass and taking into account  the observed scatter.

 Results for the NIHAO sample are overplotted as orange triangles to the 
observational data in Fig. \ref{fig:mstarrlast}: by design they occupy the same region of the $M_{\rm star}-R_{\rm last}$ plane as observed galaxies.   
 Figure \ref{fig:mstarrlast} further indicates that both NIHAO and SPARC + LITTLE THINGS  datasets cover homogeneously the $10^7-10^{11}$ M$_\odot$ stellar mass range.

\begin{figure*}
\hspace{0.cm}\includegraphics[width=\linewidth]{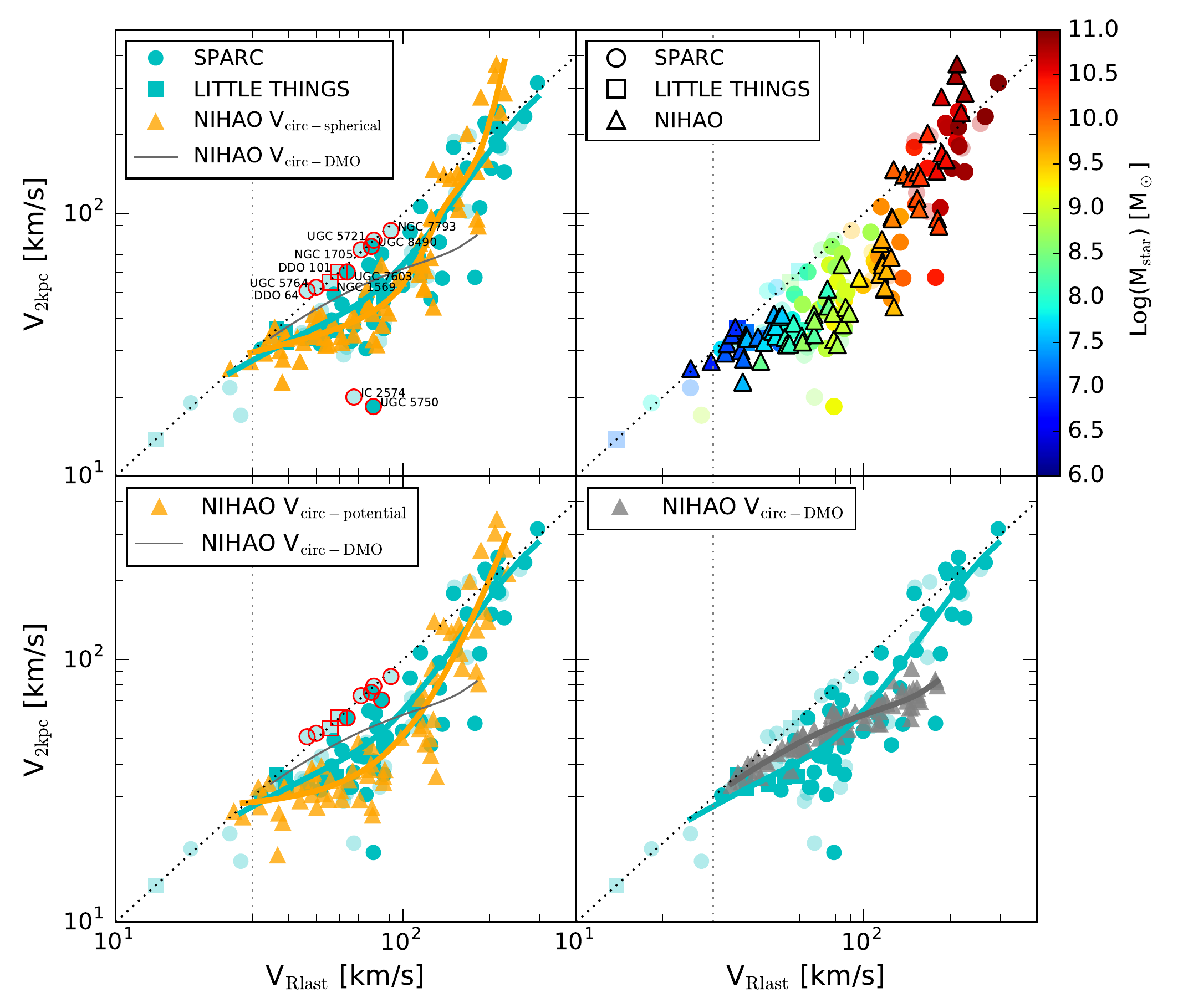}
\caption{$V_{\rm 2kpc}-V_{\rm Rlast}$ relation for SPARC (circles),
LITTLE THINGS (squares)
 and NIHAO (triangles) galaxies.  Semi-transparent 
datapoints
  indicate 
low-quality RC data according to the SPARC database and to the discussion in \citet{Iorio17}.
 NIHAO galaxies  show a large scatter in shapes, covering a wide range of $V_{\rm 2kpc}$ values for a given $V_{\rm Rlast}$ and matching in general 
observed 
  results.
\textit{Top left panel}: NIHAO shows spherical circular velocities.
Solid lines show the spline interpolation to 
observational
 and NIHAO data points, highlighting how the two distributions follow similar trends in the  $V_{\rm 2kpc}-V_{\rm Rlast}$ plane.
Discrepancies are found in the V$_{\rm Rlast}\sim 50-90$ \kms, region where NIHAO does not reach the highest and lowest $V_{\rm 2kpc}$ values observed. 
Highlighted with a red contour 
  are those observed galaxies whose $V_{\rm 2kpc}$  are outside the $\pm 3\sigma$ scatter from the expected mean NIHAO $V_{\rm 2kpc}$ value. Such observational outliers are analyzed in \Sec{notmatched}.
\textit{Top right panel}: Same as left panel, with each galaxy now colored by its stellar mass, highlighting the expected correlation between $V_{\rm Rlast}$ and $M_{\rm star}$. The stellar mass range at which the highest diversity in RCs is observed, $M_{\rm star}\sim$10$^{8.0-9.0}$\msun, corresponds to the region of maximal halo expansion of NIHAO galaxies \citep{Tollet16}, in line with results from other sets of hydro-simulations \citep[e.g.,][]{Governato12,DiCintio14a,chan15}.
\textit{Bottom left panel}: Same as top left panel, with NIHAO showing the true circular velocities, computed from the gravitational potential on the disc plane.
\textit{Bottom right panel}: Same as top left panel, with NIHAO showing the DM-only circular velocities.
The DM-only NIHAO average trend is shown for comparison in the rest of panels as a thin black line.}

\label{fig:v2vrlast}
\end{figure*} 

\section{Results}  \label{sec:results}
\subsection{Matching the diversity of observed RC shapes} \label{matching}

To compare the global shapes of simulated and observed galaxies' RCs we use the $V_{\rm 2kpc}-V_{\rm Rlast}$ relation,  where $V_{\rm 2kpc}$ is the RC at 2 kpc from the centre, and $V_{\rm Rlast}$ is the RC measured at $R_{\rm last}$, given by Eq. \ref{eqrlastmstar} for NIHAO galaxies. 
For those observed  galaxies that do not have measured velocities exactly at 2 kpc, 
we interpolate between 
the two closest bins.
 This relation is shown in Fig. \ref{fig:v2vrlast},
where in all four panels SPARC data are shown as circles, 
LITTLE THINGS as squares, and
 NIHAO galaxies as triangles.

The top left panel of Fig. \ref{fig:v2vrlast} shows the $V_{\rm 2kpc}-V_{\rm Rlast}$ relation of observed (cyan) and NIHAO (orange) galaxies, 
where NIHAO rotational velocities are 
approximated as the spherical circular velocities: 
$ V_{\rm circ-spherical}(r)=\sqrt{\frac{G M(<r)}{r}} $,
where $r$ is the 3D radius and $M(<r)$ is the total mass enclosed within such radius.\footnote{
We acknowledge that assuming a spherical mass distribution
is not the accurate-most hypothesis,
 however we use this definition in order  to be able to
directly compare our results to previous simulation data that use it as well: in particular the EAGLE/APOSTLE simulations presented in \citet{Oman15}.
}
SPARC galaxies with high quality H\,{\sc i}-H~$\alpha$ RC measurements 
(as evaluated in the SPARC dataset)
are indicated with dark cyan circles, while lower quality data are shown as semi-transparent cyan circles.
The lower quality of H\,{\sc i} data in these galaxies is due to major asymmetries, strong non-circular motions, and/or offsets between the H\,{\sc i} and stellar distributions.
%
%
Similarly, LITTLE THINGS galaxies with very uncertain RCs 
\citep[due to peculiar kinematics, strong asymmetries, or poorly determined distances, as discussed in][] {Iorio17}
are also shown with semi-transparent color.
Additionally, this panel shows as solid lines the smoothed-spline interpolations to observational and NIHAO data,  highlighting
  the average distribution of observed and simulated galaxies in the $V_{\rm 2kpc}-V_{\rm Rlast}$ plane.

From the top left panel of Fig. \ref{fig:v2vrlast} we see that the simulated NIHAO galaxies
 occupy a similar region of the $V_{\rm 2kpc}-V_{\rm Rlast}$ plane
and follow
 similar trends as the observed galaxies.
Furthermore, they show a wide variety of RC shapes for a given $V_{\rm Rlast}$.
 This is a noticeable improvement compared to the hydro-simulations analyzed in \citet{Oman15,Oman16} that instead follow  the expectations of a DM-only model  with very little scatter up to $V_{\rm max}\sim$30 \kms 
 (as the 1:1 dotted line), and then start deviating from it at $V_{\rm max}$
 larger than 60--70 \kms 
  (larger halo masses) due to the increasing contribution of baryons at the galactic centre.

The top right panel of Fig. \ref{fig:v2vrlast} shows SPARC (circles), LITTLE THINGS (squares), and NIHAO (triangles) data colored by stellar mass, again with less accurate observational RCs indicated with a semi-transparent color. As expected, there is a correlation between $V_{\rm Rlast}$ and $M_{\rm star}$. 
This panel highlights that both observed and simulated galaxies with similar stellar masses occupy similar positions in the $V_{\rm 2kpc}-V_{\rm Rlast}$ plane.
 It also shows that the region at which the spread in RC shapes is maximal, i.e.,  between 50$<V_{\rm Rlast}$\kms$<$100, corresponds to the 
 stellar-to-halo mass ratios
 at which we expect DM core formation to be most efficient:  
after assuming an abundance-matching relation this corresponds to 
$M_{\rm star}$ between $10^7$ and $10^9$ M$_\odot$ \citep{DiCintio14a}.  
  Therefore, the diversity of RC shapes highlighted by \citet{Oman15}  primarily occurs at the precise mass range where the models show the greatest effects  of baryonic physics on
 modifying the density profiles of cold-DM halos.

Given that assuming a spherical mass distribution is a simplification, we show in 
the bottom left panel of Fig. \ref{fig:v2vrlast}  the `true' circular velocities of NIHAO galaxies, computed through the gravitational potential in the disc plane:  
$V_{\rm circ-potential}= \sqrt{R \frac{\partial \Phi}{\partial R}}_{\rm z=0}$,
  such that the actual mass distruibution of disc galaxies is taken into account.
  Note that
  the H\,{\sc i} gas rotational velocity traces the  true circular velocity of the galaxy if its gas is in equilibrium \citep{BinneyTremaine08,Burkert10,Iorio17}.
 Though this may not be the case of all simulated (or observed) galaxies studied here,
the use of the true circular velocity allows a closer comparison
with H\,{\sc i} RC observational data.

When using $V_{\rm circ-potential}$,
 the overall scatter in the
  $V_{\rm 2kpc}-V_{\rm Rlast}$ plane  is 
slightly  
  increased 
and the mean trend line is
 lowered.
 This  lower average $V_{\rm 2kpc}$ result is expected since
 we recall that,
 in the extreme case of an exponential disc mass distribution, 
  $V_{\rm circ}$ increases  linearly with $R$ at radii smaller than the scale-length, 
instead of as 
$\sqrt{R}$  as in the spherical case.
This inner region discrepancy can be further enhanced if there is an H\,{\sc i} hole in the galactic centre, as is commonly observed. 
More specifically, we find  
$\rm \frac{V_{\rm 2kpc}^{ circ-potential}}{V_{\rm 2kpc}^{circ-spherical}} > 0.76 $,  
 with a mean of 0.91, and 
 $\rm \frac{V_{\rm Rlast}^{ circ-potential}}{V_{\rm Rlast}^{circ-spherical}} \approx 0.99$.
Thus, these not very remarkable quantitative differences between the two definitions of circular velocity 
 allow for  the same qualitative analysis and general conclusions formerly drawn for the spherical circular velocity case.

 A more thoughtful comparison between simulated and observed galaxies, however, requires refined methods in order to trace the actual dynamics of the gas: indeed, it has been shown by several authors  
 that the H\,{\sc i} velocity profile of the gas (V$_{\rm HI}$) in dwarf galaxies often does not trace the circular velocity of the corresponding halo \citep{Valenzuela07, brookdicintio15, Maccio16, Pineda17, Verbeke17, ElBadry17b, Brooks17, Oman17}. 
As already mentioned, 
  in this case new factors such as equilibrium criteria, velocity dispersions, and asymmetric drift corrections will come into play, making the comparison more robust yet more complicated. These issues will be studied in a forthcoming paper (Dutton et al. in prep).

Finally, 
the bottom right panel of Fig. \ref{fig:v2vrlast} shows NIHAO's DM-only circular velocities.
Without hydrodynamics, NIHAO  DM halos are self-similar, which results in  self-similar circular RCs and minimum scatter on the $V_{\rm 2kpc}-V_{\rm Rlast}$ relation.
 This panel stresses how much pure \lcdm\ predictions are at odds with observational RC data, and how baryonic effects on DM are crucial to reproduce observed galaxy properties.
The DM-only NIHAO prediction is placed as a thin black line on the rest of panels for a clearer comparison with the cases where baryons are included.

  \vspace{0.5cm}
  
 The  success of NIHAO hydro-simulations in recovering the observed distribution and scatter of RC shapes 
 lies in  the wide range of halo responses these simulated galaxies can exhibit, from contraction to expansion.
 In particular,
 the match with data
  at the controversial scales of intermediate-mass and dwarf galaxies, is
 due to the role of supernova-driven gas outflows that induce the expansion of the central galactic DM  content \citep{Governato12,pontzen12,Maccio12}.
These processes generate a DM (and stellar) core in the galaxy, in a mass-dependent
 fashion
 \citep{DiCintio14a,Tollet16,DiCintio17}:  the core formation mechanism is strongly sensitive to even small  variations in $M_{\rm star}$   \citep[as already reported in][]{brook15a};
and at a given stellar mass, there is in addition a tendency for 
 larger galaxies to expand more \citep{Dutton16}. 
All this results in NIHAO achieving a relatively wide variety of RC shapes.
 Furthermore, the mass-dependent core formation model presented in \citet{DiCintio14b} predicts that at the lowest
   mass scales,
    i.e., for galaxies with $M_{\rm star}<10^6$ \msun,
     the effect of stellar feedback is 
     insufficient
      to create a shallow DM  distribution \citep{Penarrubia12,Governato12}.
   This should translate into a similarity of RCs at this extreme mass scale, which is indeed evident both in observations and simulations with $V_{\rm Rlast}\lesssim 40$ \kms.
%
At the highest mass scales,
 i.e. $M_{\rm star}>10^{10.5}$ \msun,
  the effect of stellar feedback  starts to be insufficient to modify the DM halo 
  due to the deepening of the potential caused by stars that form in the central region.
   However,
   precisely
  because of this increasing dominance of baryons at the galactic centre,
  their RC shapes are not expected to be self-similar but to vary, 
 tracing the different possible final baryonic distributions.
  This explains the existing scatter in both observed and simulated RC shapes above $V_{\rm Rlast}\gtrsim 150$ \kms.

Due to the choice of the quantities plotted,
$V_{\rm 2kpc}$ will be equal to $V_{\rm Rlast}$ for observed and simulated galaxies with $R_{\rm last}=2$ kpc. According to Eq. \ref{eqrlastmstar} and to the top-right panel of Fig. \ref{fig:v2vrlast}, this corresponds to a typical value of $V_{\rm Rlast}\sim 30$ \kms,
marked in all panels with a thin dotted vertical line.
 We insist however that the choice of using the velocity at a radius of 2 kpc does not force the match  at the low-velocity end between observational and simulation results. Figure \ref{fig:newv1kpc} shows that when plotting instead the velocity at a smaller radius, for example at 1 kpc,
the least massive galaxies also tend to have higher $V_{\rm 2kpc}$, coming closer to the 1:1 line, and in general the shapes of the trendlines of both the observed and simulated samples coincide. 

The main difference with respect to the $V_{\rm 2kpc}$   figure is that with $V_{\rm 1kpc}$ there is a higher dispersion over the entire $V_{\rm Rlast}$  range, both in the observational data and in the simulation results.
We note that when using instead $V_{\rm 3kpc}$,  the average trend keeps the same curved shape but this time with a smaller dispersion of both simulated and observed results.
Therefore, results with velocities computed at 1 kpc, 2 kpc, and 3 kpc show qualitatively the same trends, but with a decreasing overall scatter as we move to larger radii.
The fact that the overall scatter in the $V_{\rm fixed\,r[kpc]}-V_{\rm Rlast}$  plane decreases when computing velocities at higher and higher radii
 is an interesting point: this is what we expect from models in which core formation mainly acts at modifying the inner region of DM halos, such as baryonic-driven core formation. As we move towards large radii, $R>3$ kpc, the effect of core formation is less evident on the RC of galaxies, and therefore we expect to observe less scatter in their RC shapes. This is indeed the case for both NIHAO simulations and the observational results presented here.

 \begin{figure}
\hspace{0.cm}\includegraphics[width=\linewidth]{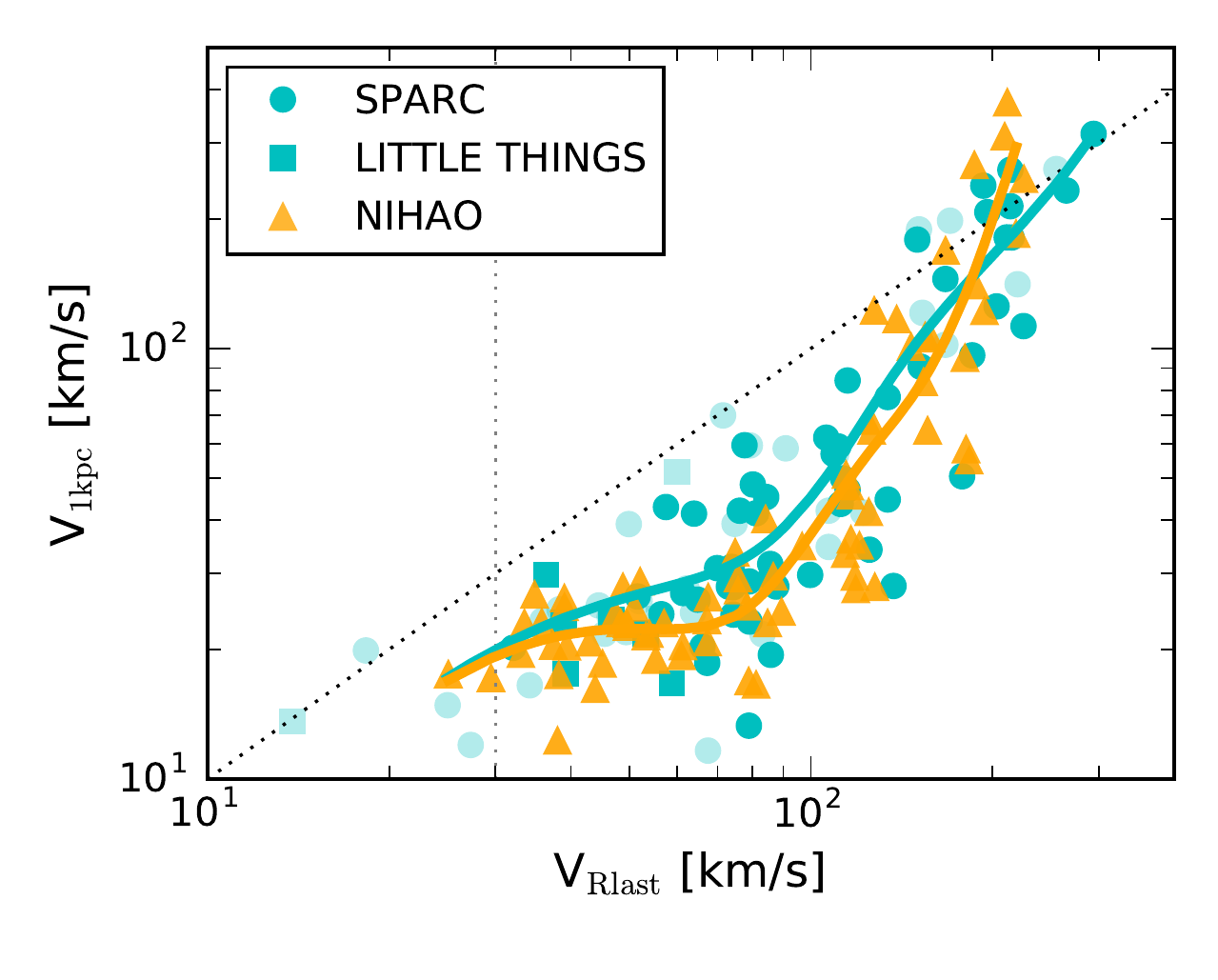}
\caption{
The $V_{\rm 1kpc}-V_{\rm Rlast}$ relation. The match of the observed and simulated average trends at the low-velocity end also occurs when  plotting the velocity at 1 kpc instead of at 2 kpc.  
}
\label{fig:newv1kpc}
\end{figure}

  \vspace{0.5cm}

 Although the NIHAO simulated galaxies  follow the observational trend quite well, the match to observational data is not perfect. 
We denote as `outliers' those observed galaxies whose $V_{\rm 2kpc}$ is outside the $\pm 3\sigma$ range determined with respect to the NIHAO trend line.
Note that we do not take into account the individual measurement errors when checking if observational data are inside the $\pm 3\sigma$  contours. 
For SPARC these are, on average, between $5-10$ \kms, and
include the uncertainties in circular random motions and distance estimates but do not include the systematic inclination error. 
The LITTLE THINGS errors are of $\sim 5$ \kms\ on average and account for the errors coming from the $^{\rm 3D}$BAROLO parameter estimations and the asymmetric drift corrections.
If the error bars are considered, some of the following galaxies are actually not outliers to NIHAO. We stick here however to this more conservative approach.
  

 Focusing on the top left panel of Fig. \ref{fig:v2vrlast}, where spherical circular velocities were used for NIHAO,
we find 
11
 observational outliers, all between   
$47<V_{\rm Rlast}$/\kms$<90$, 
  which we circle in red: they either present an extremely steep RC (falling close to the 1:1 dotted line), or an extremely shallow one.
These galaxies will be studied  in detail in \Sec{notmatched}.
When using instead the true circular velocities,
 we find 1 additional outlier with a steeply rising RC, but no longer the shallow rising galaxies (see red circles in bottom left panel of Fig. \ref{fig:v2vrlast}). In fact, the RC of these galaxies are matched very well by NIHAO galaxies of similar $V_{\rm Rlast}$ (see Sec. \Sec{notmatched}).
It is worth mentioning that the steepest observed RCs in the $\sim 50-100$ \kms\ range pose a problem also to simulations that predict a cuspy inner mass profile:  when put on the $V_{\rm 2kpc}-V_{\rm Rlast}$ plane, the EAGLE/APOSTLE simulations are also unable to match the SPARC outliers with the highest $V_{\rm 2kpc}$
velocities, despite the high mean  $V_{\rm 2kpc}/V_{\rm Rlast}$ values they present
over the whole $V_{\rm Rlast}$ extent
(see Fig. \ref{fig:binned}).

We  note that at the highest $V_{\rm Rlast}$ values ($M_{\rm star}\gtrsim 10^{10.7}$ M$_\odot$) some NIHAO RCs are rising more steeply than SPARC galaxies. 
NIHAO simulations do not include the effects of  AGN feedback, which is particularly important at these mass scales to counteract the cooling of baryons  and regulate  star formation \citep{Nulsen07,McCarthy10,Beckmann17}.
Therefore they may enclose a higher-than-average mass in their inner regions compared to SPARC data.
The significance of the deviation of simulated RCs from observed ones at the highest stellar masses  will  be explored in future work that takes into account  AGN effects.

We caution the reader that while the observations (SPARC + LITTLE THINGS) and the simulations (NIHAO) are representative, they are not necessarily unbiased samples of the underlying galaxy population.  Thus, in the
$V_{\rm 2kpc}-V_{\rm Rlast}$
 plot the differences between various simulations and observations  may be due in part to sampling effects, and this issue applies as well  to all previous studies on the subject.

\vspace{0.5cm}

Finally, in Fig. \ref{fig:binned}
we quantify the scatter of the $V_{\rm 2kpc}-V_{\rm Rlast}$ relation by dividing the data in regular logarithmic bins in V$_{\rm Rlast}$ and comparing the simulated and observed mean  $V_{\rm 2kpc}/V_{\rm Rlast}$ ratios and standard deviations in each bin. 
As in previous figures, observational data is shown in cyan and NIHAO results (note that these are the spherical circular velocities) 
in orange.
 %
The mean values are given by the interpolation lines in
the top left panel of
 Fig. \ref{fig:v2vrlast},
while the scatter indicates the standard deviation to this mean $V_{\rm 2kpc}/V_{\rm Rlast}$ ratio, thus representing  a direct measure of the diversity in RC shapes.

The mean $V_{\rm 2kpc}/V_{\rm Rlast}$ values are reasonably similar  in the observational and NIHAO samples,
and
the major differences in the size of scatter appear again in the previously mentioned regions:
 in the high $V_{\rm Rlast}$ range where 
NIHAO RCs rise steeper than SPARC (probably due to the lack of AGN effects),
  and in the $V_{\rm Rlast}\sim$50--100 \kms\ range 
 where there are observed galaxies with both steeper and shallower RCs than NIHAO (outliers 
 with red contours in top left panel of Fig. \ref{fig:v2vrlast}). We note that when using the true circular velocities, the overall scatter  is increased 14 per cent on average, and the mean $V_{\rm 2kpc}/V_{\rm Rlast}$ ratios are decreased 7 per cent on average.

Figure  \ref{fig:binned} also shows the results  for the EAGLE/APOSTLE simulations from \citet{Oman15} (red squares).
Their $V_{\rm Rlast}$ values as calculated from Eq. \ref{eqrlastmstar} have been kindly provided by Kyle Oman.
%
Since these simulations form galaxies with self-similar cuspy DM halos at every mass, their mean $V_{\rm 2kpc}/V_{\rm Rlast}$ values  in the  $V_{\rm Rlast}\sim 50-100$ \kms\  region are significantly higher than observed, while their measured scatter is  smaller than observed. 
Thus, the greatest mismatch 
between EAGLE/APOSTLE simulations  and observations
 occurs 
in the precise $V_{\rm Rlast}$ range where the NIHAO galaxies are most efficient at forming cores.
 As compared to NIHAO data, EAGLE/APOSTLE  $V_{\rm 2kpc}/V_{\rm Rlast}$ scatter is on average 60 per cent that of NIHAO.  

\begin{figure}
\hspace{0.cm}\includegraphics[width=\linewidth]{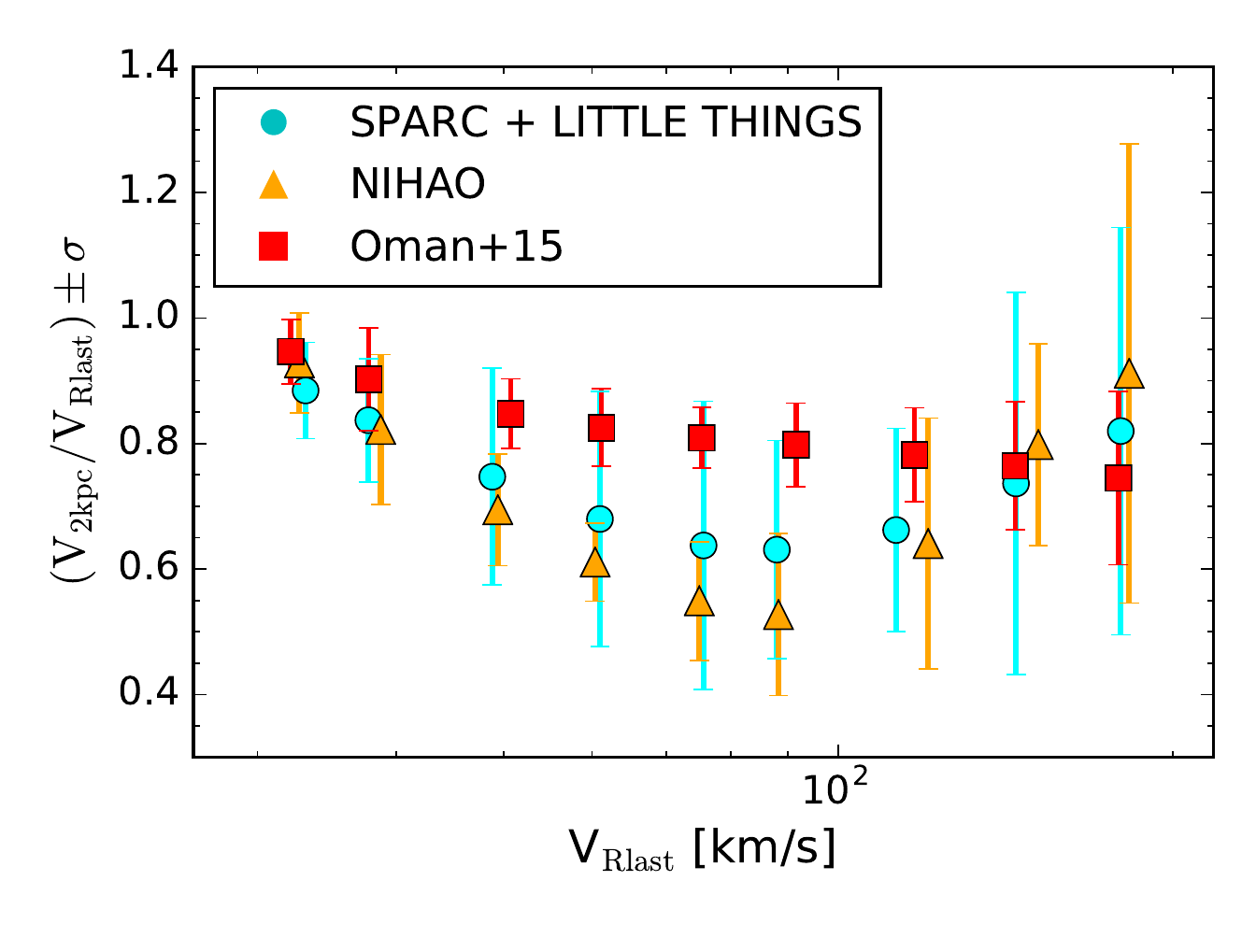}
\caption{The $V_{\rm 2kpc}/V_{\rm Rlast}$--$V_{\rm Rlast}$ relation, binned by $V_{\rm Rlast}$, of observed galaxies (SPARC + LITTLE THINGS) and  NIHAO and EAGLE/APOSTLE simulations.
NIHAO and EAGLE/APOSTLE show the spherical circular velocities (see top-left panel of Fig. \ref{fig:v2vrlast} for NIHAO).
  Points show the mean $V_{\rm 2kpc}$/ $V_{\rm Rlast}$ ratios given by the interpolation lines in Fig. \ref{fig:v2vrlast}.
The scatter is calculated as the standard deviation of the $V_{\rm 2kpc}$/ $V_{\rm Rlast}$ ratio to its mean expected value in each bin, representing a measure of the diversity in RC shapes.}
\label{fig:binned}
\end{figure}

\subsection{Observational outliers} \label{notmatched}

We  analyze in detail the individual galaxies from the observed samples that are outliers in the $V_{\rm 2kpc}-V_{\rm Rlast}$ plane, highlighted with red contours in Fig. \ref{fig:v2vrlast} (top left panel).
They correspond to  galaxies whose velocity at 2 kpc is more than $\pm$3$\sigma$ away from  the mean NIHAO value in the same $V_{\rm Rlast}$ range.
As said previously, this is a subset of
 11
  galaxies, all within $47-90$ \kms\ in $V_{\rm Rlast}$.
We recall that we have not taken into account  the measurement error for each observed galaxy when assessing if they are $\pm$3$\sigma$ away from simulation averages.

The RCs of these outlier galaxies are depicted with colored lines in Fig. \ref{fig:rcnot} with their names indicated in the legend. The R-axis is truncated at 12 kpc for clarity, but we note that galaxy UGC~5750 extends flat out  to 23 kpc.
This subset of galaxies is compared to the sample of NIHAO galaxies within the same  $V_{\rm Rlast}$ range, $47-90$ \kms,  shown as grey lines in Fig. \ref{fig:rcnot}.
All the NIHAO galaxies in this range form cores of various sizes,  matching very well the remaining  
31
 observed galaxies in this velocity region which represent the majority of observed SPARC galaxies here 
($\sim74$ per cent).

Apart from galaxies IC 2574 and UGC~5750, which have a very shallow RC, all the remaining
9
 outliers  do not match NIHAO expectations since their RCs are steeper, 
rising faster from the galactic centre, 
and presenting a sharp discontinuity to flat rotation.
Such extremely steep 
 RCs are too steep even compared to expectations from simulations with a universal standard NFW profile for DM halos \citep{Oman15},
and are not matched by any NIHAO DM-only RC either (see bottom right panel of Fig. \ref{fig:v2vrlast}). This may be indicating that 
i) the baryon content is not negligible compared to the DM,
or
ii) the data is not tracing the potential, probably caused by the presence of non-circular
and/or random
 motions, or fast moving H\,{\sc i} bubbles  \citep{Read16} around the galactic centre.
Further,
these galaxies are outliers compared not only  to  simulated  galaxies, but also to the average observed RCs in such $V_{\rm Rlast}$ range.

We note that out of 
11
 outliers only
3
 have high quality RC data, while the remaining 
8 
  present several observational and technical issues that hinder an accurate derivation of their RCs. 
Specifically, the sources of error in the determination of the velocity fields of these galaxies include
(for references and further details see Secs. \ref{sec:lowv2} and \ref{sec:highv2}):
 disturbed H\,{\sc i} discs containing holes or being severely warped
 (e.g., NGC~7793, NGC~1705,
  UGC~5764, IC~2574, NGC 1569),
  significant uncertainties in distances, inclinations
and orientation angles  
   (e.g., UGC~5721,
    NGC~7793, DDO~101), abrupt changes of inclination throughout the galaxy disc 
 (e.g., NGC~1705),
  unsufficient angular resolution over the beam (e.g., 
  UGC~5721), 
   asymmetries and non-circular or random motions (e.g., IC~2574, DDO~64, NGC 1569).

We firstly note that with such observational uncertainties it seems natural that the scatter in observed RC shapes should be larger than that found in simulations. 
Secondly, we note that some environmental effects may be missing in the NIHAO sample, which are simulations of  isolated field galaxies only. The SPARC and LITTLE THINGS datasets, instead, 
include galaxies both in isolation and within groups. Such a difference 
could represent an extra  source of higher scatter in observations, since a group environment can strongly affect a galaxy's evolution \citep[see e.g.,][]{Wetzel13,DelPopolo12}.

We proceed by examining one by one the
 outliers shown in the top-left panel of Fig. \ref{fig:rcnot}.


\subsection*{Low-V$_{\rm 2kpc}$ outliers} \label{sec:lowv2}
 

 Galaxies  IC~2574 and UGC~5750
  have been extensively studied elsewhere in the context of \lcdm\ \citep[e.g.,][]{Oman15,Oman16,Creasey17}:
 their RCs are so slowly-rising that their implied DM core is even larger than that predicted by other hydro-simulations that  form cores. While issues about non-circular motions have been raised by \cite{DeBlok08} and \cite{oh08} for the case of IC~2574, such concerns do not apply to  UGC~5750.  
Therefore, it has been argued that these galaxies pose a problem to the core-formation scenario \citep{Oman15}. We show here that this is not the case.

In Fig. \ref{fig:coreRC} we closely compare  the observed RC  of
IC~2574 and UGC~5750
 with 
that of their most-similar NIHAO counterpart
(g9.59e10 and g1.59e10, respectively).
SPARC data is presented with error bars.
 For NIHAO galaxies we show 
 the circular velocities from the DM-only run $V_{\rm circ\, DMO}$,
 the spherical circular velocities 
 $V_{\rm circ-spherical}$ (as in top left panel of Fig. \ref{fig:v2vrlast}),
  and the true circular velocties computed from the gradient of the gravitational potential at the disc plane
   $V_{\rm circ-potential}$ (as in bottom left panel of Fig.    \ref{fig:v2vrlast}).

 Although 
IC~2574 and UGC~5750
  present a higher than 3$\sigma$ difference with respect to NIHAO galaxies on the $V_{\rm 2kpc}-V_{\rm Rlast}$ plane
when using  $V_{\rm circ-spherical}$, 
the use of the true circular velocity that takes into account the precise mass distribution of the galaxy  provides a  remarkable agreement with observational data,  proving that when simulations include effective feedback, dwarf galaxies IC~2574 and UGC~5750 
can be reconciled with predictions based on the \lcdm\ model.
We note that in the case of these two particular NIHAO galaxies, gas velocity dispersions are low and asymmetric drift corrections can be neglected, which
 means that
this   $V_{\rm circ-potential}$ is approximately equal to the rotational velocity that could be measured from the H\,{\sc i} gas.
 This may not be the situation for all simulated  NIHAO galaxies; therefore, as said previously, 
this
 more realistic comparison
of the H\,{\sc i} gas velocities of NIHAO simulations with observations 
  is left for a following paper.
We already hint here that, as presumed, adding observational `effects' or `biases' to the derivation of simulated galaxies' RCs will provide very diverse RC shapes and add a considerable amount of scatter to the  $V_{\rm 2kpc}-V_{\rm Rlast}$ relation yielding a better match with observational data.

\begin{figure}
\hspace{0.cm}\includegraphics[width=\linewidth]{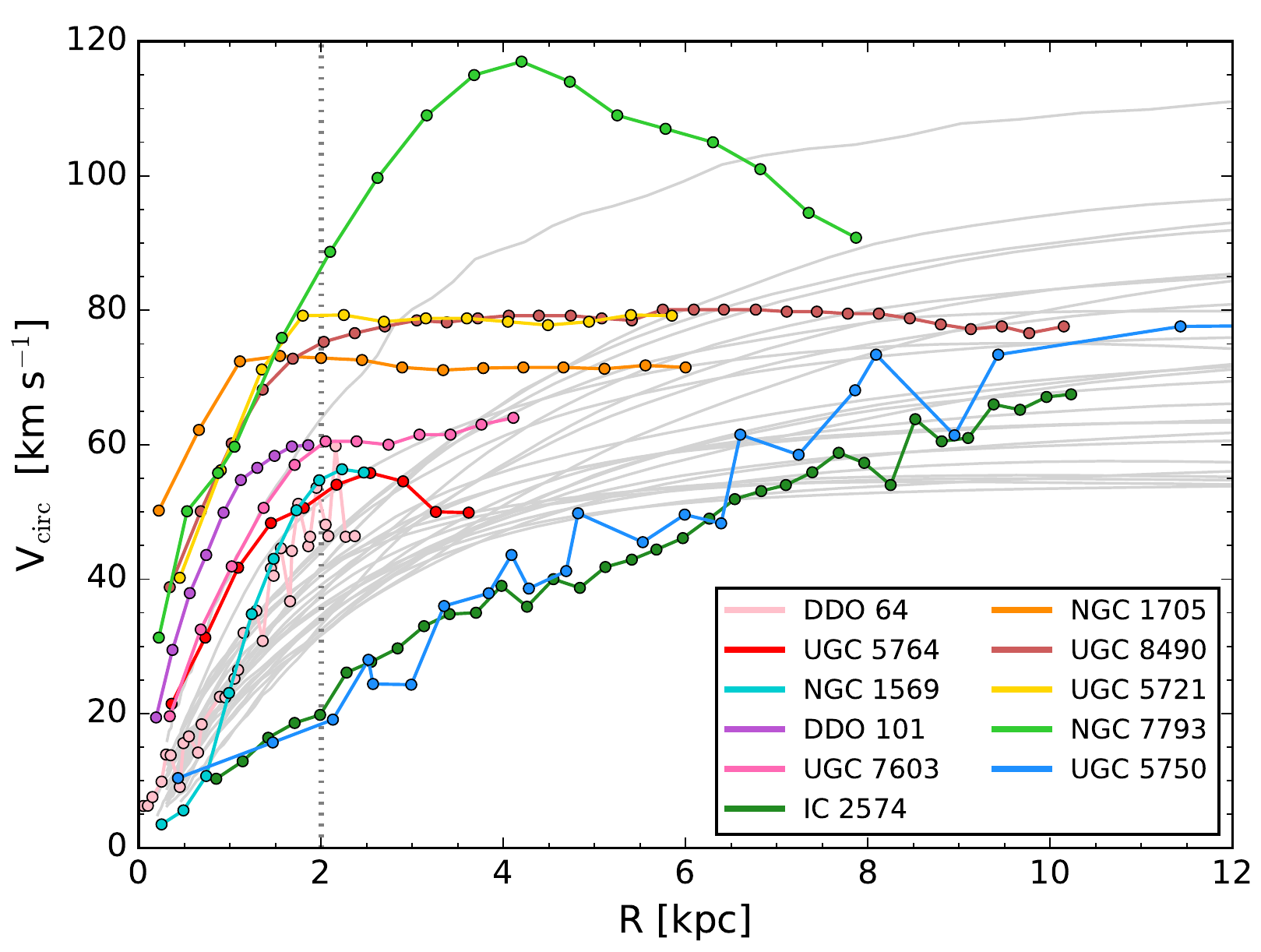}
\caption{RCs of  NIHAO galaxies with $47<V_{\rm Rlast}/$\kms$<90$  (gray lines), compared to observed outlier galaxies within the same V$_{\rm Rlast}$ range  (colored lines).
These outliers are those observed galaxies  whose $V_{\rm 2kpc}$ is  $\pm$3$\sigma$ further away from the average  NIHAO $V_{\rm 2kpc}$ value,  
  corresponding to the galaxies with red contours in
the top-left panel of  
   Fig. \ref{fig:v2vrlast}.  
A dotted line marks the velocity at 2 kpc.
 }
\label{fig:rcnot}
\end{figure}

\begin{figure*}
\hspace{0.cm}\includegraphics[width=\linewidth]{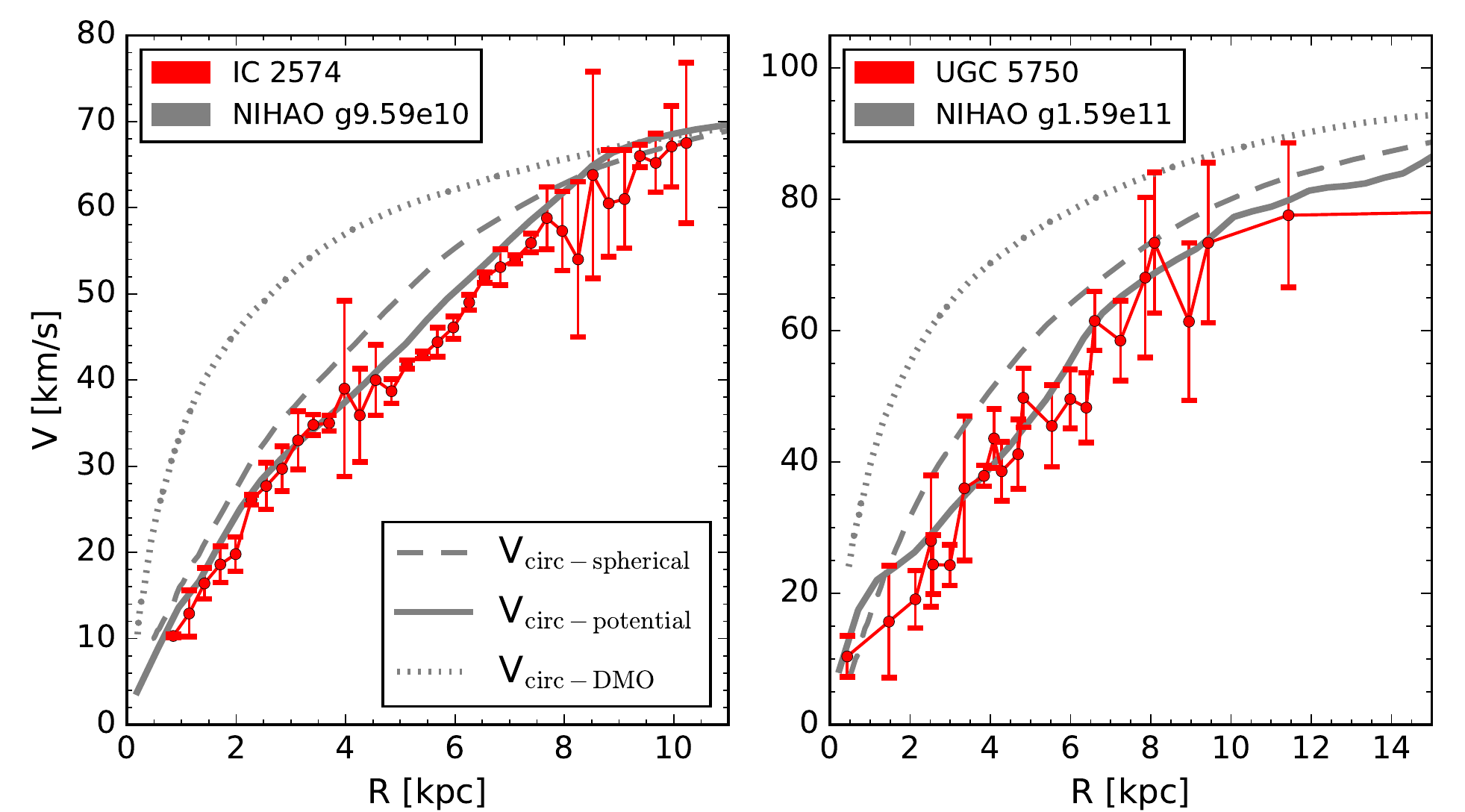}
\caption{Rotation curves of galaxies IC~2574 and UGC~5750 compared to that of their most similar NIHAO counterpart. For NIHAO galaxies we show the DM-only circular velocity, the spherical circular velocity, and the true circular velocity, computed through the gravitational potential on the disc plane. In this last case, the agreement with observed data is remarkable, showing that
the formation of extremely cored dwarf galaxies is indeed possible within  a \lcdm\ cosmology, and highlighting the importance of properly
comparing observational and theoretical velocities.
}
\label{fig:coreRC}
\end{figure*}

\subsection*{High-V$_{\rm 2kpc}$ outliers}   \label{sec:highv2}

Galaxy NGC~1705  has  a regularly rotating H\,{\sc i} disc which is strongly warped, and may be out of equilibrium due to its off-centred stellar component. Moreover, its H\,{\sc i} disc may indicate the presence of radial motions (15 \kms)  near the centre \citep{Lelli14b}.   NGC~1705 is a starburst galaxy \citep{Annibali03}, with a birthrate parameter close to 6
\citep{Lelli14b},
 where the \citet{McQuinn09} definition of  starburst is followed, as a galaxy that has  a birthrate parameter  $\rm b=SFR(last 1 Gyr)/SFR(mean)_{0-6} > 3$, where SFR(mean)$_{0-6}$ is the mean SFR over the past 6 Gyr.

UGC~8490
(also known as NGC~5204)
  is a 
 nearby compact irregular
 galaxy that is strongly warped, with several changes in inclination \citep{Sicotte97}. The innermost points of its H~$\alpha$ and H\,{\sc i} RCs do not agree \citep{swaters09}.
 Its star formation history shows a significant rise at recent times compared with the average star formation rate over the last few Gyr,   in consistency with known starburst dwarf galaxies 
 \citep{Lelli14a,McQuinn15}.

UGC~5721 (NGC~3274) is a barred spiral, emission-line or 
H\,{\sc ii}
 galaxy  \citep{Hunter82} in the constellation of Leo.
Its H~$\alpha$ curve  shows a steep rise  that is not seen in the H\,{\sc i} data. Moreover, the inner points of its RC are uncertain because of insufficient angular resolution
\citep{swaters09}. It has a circular velocity gradient comparable to those of BCDs \citep{Lelli14a}, and so could  possibly harbor a starburst.

UGC~7603 (NGC~4455) is another emission-line (Seyfert) galaxy \citep{Deo07}.
The derived H\,{\sc i} RC may be uncertain since its velocity gradient is not well resolved by the H\,{\sc i} beam.
 This uncertainty is enhanced in the inner region of the RC due to its high inclination
 \citep{swaters09}.


UGC~5764 (DDO 83) 
is a gas-rich isolated galaxy, classified morphologically as irregular or even BCD. It has a low luminosity but high central surface brightness giving a $<1.5$ kpc scale length.
 This galaxy seems to have just started to flatten out at the edge of its detected H\,{\sc i}, but it
   presents a warp on its receding side that makes its last 2 points uncertain \citep{VanZee97}.

NGC~7793 is a warped spiral galaxy in the constellation of Sculptor, at a distance of 
$\sim 4$ Mpc.
It  has a chaotic spiral structure and shows signs of new star formation. Its HI RC appears to be declining in the outer parts \citep{Carignan90,Dicaire08,DeBlok08} though the inclination estimates are uncertain. It hosts many X-ray sources; e.g., in the outskirts lies a powerful microquasar that is driving gas away from the disc at high speeds in two jets, creating a 300-parsec-long jet-inflated bubble of hot gas  \citep{Pakull10}.


NGC 1569 (UGC 3056) is a nearby ($\sim 2-3$ Mpc) starburst dwarf irregular galaxy with very disturbed HI kinematics and a turbulent ISM. Its velocity dispersion is $\sim 20$ \kms\ on average and therefore the asymmetric drift correction dominates the RC at all radii \citep{Iorio17}.
Several young super stellar clusters are located in its centre. It also presents signs of superbubbles and supergalactic winds \citep{SanchezCruces15}.

DDO~64 (UGC~5272) is a dwarf irregular galaxy that is actively forming stars. It is dominated by HII regions around which stars are clumped, giving place to a patchy distribution. The inner regions of the stellar component are dominated by a young population which translates in this galaxy presenting extremely blue colors as compared to other irregular galaxies. Emission lines are measured from spectroscopy of its HII regions \citep{Hopp91}. 
Non-circular motions may be causing a difference between the approaching and receding sides of the velocity field \citep{deblok02}.

DDO 101's HI disc is extended only slightly beyond the optical disc. It's RC only reaches out to a radius of $R=1.9$ kpc \citep{Iorio17}. The estimates of the distance for this galaxy are very uncertain, ranging from 5 to 16 Mpc.

So, of the 
9
galaxies with steeply rising RCs, 
7 are either starburst galaxies or emission-line galaxies, which may also be indicative of significant ongoing star formation, 
since the gas is presumably
ionized by the ultraviolet radiation from massive young stars.
 Indeed, that the inner RC of starburst dwarfs rises more steeply than that of typical dwarf irregulars has been shown in previous studies \citep{Lelli14c,Lelli14b,McQuinn15},
indicating
that starburst galaxies have a higher-than-average central dynamical mass and that  starburst activity may be related to the shape of the potential well.

In \citet{Lelli14b} it is argued that the steep RC of starburst galaxies is in disagreement with the `cusp-core' transformation picture offered by cosmological hydrodynamical simulations.
These galaxies observationally show a large gas reservoir which the authors discuss could drive large fluctuations of the central potential forming eventually  a shallow core.
 This mechanism is particularly efficient during violent star formation episodes, such as those a starburst galaxy is going through. However, the timescale and total duration of such stellar activity should also be taken into account. Models of DM core creation, based on cosmological simulations, agree on the notion that a bursty, repeated, and \textit{extended} star formation history is necessary in order to create a large DM core \citep[e.g.,][]{pontzen12, Read16a, DiCintio17}. 
By contrast,
star formation in starburst galaxies commonly happens at very late times,  after a possibly overall quiet star formation history.
If proven by future observations, this would support a scenario in which 
low mass
starbursts 
have a highly concentrated dark matter inner mass distribution, matched by an NFW or a contracted profile,
since there has not been enough energy from gas outflows at any time in order to modify 
the inner region.
%
Once again, note that indeed these galaxies have a central mass distribution at 2 kpc which is even higher than that of the EAGLE/APOSTLE simulations and NIHAO DM-only predictions: this highlights the role of violent and ongoing starbursts in creating a contracted central density.
That starburst phases lead to 
steep RCs, as measured from gas kinematics,
 has also been recently shown with  high resolution dwarf galaxy simulations \citep{Read16}.

In favor of this possibility is the fact that these star-forming (either starburst of emission-line) galaxies have smaller effective radii than average galaxies of the same stellar mass.
 For example, NGC~1705, UGC~8490, UGC~5721, and UGC~7603 have a $r_{\rm eff}\leqslant$ 1 kpc \citep[see][]{Lelli16}.
  In this sense, these galaxies are outliers within the observational  sample itself.
  These relatively compact stellar populations  may be indicative of  these galaxies having undergone less  halo expansion \citep[e.g.,][]{Teyssier13,Dutton16,DiCintio17}.    
%

Note that we do not claim  that
all starbursts  should have steep DM profiles.  As said previously,   the creation of a core  versus maintaining a cuspy NFW inner profile  strongly depends on the star formation activity \textit{before} the final starburst phase.
In fact,
opposed to \cite{Lelli14b}'s  general findings of
high central densities in starburst galaxies, 
 \cite{McQuinn12} showed the diverse spatial distribution of the stellar activity in a set of observed starburst galaxies, where `centrally concentrated' bursts were not the most common case.
For example, IC~2574 has a burst at $z=0$ and yet shows a large central core. 
In this case the burst occurs on the borders of a supergiant H\,{\sc i} shell located close to the edge of the galaxy \citep{Weisz09}.
We note however that
evidence of frequent 
disturbed
non-symmetric H\,{\sc i} disc morphologies observed in starburst dwarf galaxies \citep{Lelli14c}
support that the
  mechanisms  that are most likely able to  set off new star formation activity in dwarfs are  tidal perturbations from companion objects and gas shocks during mergers.
If  environmental processes are triggering star formation in these observed outliers, 
then they cannot be directly compared with the isolated galaxies of the NIHAO suite.

Of course, this picture about starburst galaxies and their relation to the `cusp-core' issue is purely speculative at this stage. Further studies based on simulations that take into account the role of environment, considering the full parameter space of interactions in a cosmological context at $z=0$,  are required to assess the origin and  properties of starburst galaxies, and their relation to the
cusp-core transformation.
Finally, more detailed  measurements of the full star formation histories of starburst galaxies are necessary in order to shed light on the interplay between star formation and the shape of DM halo profiles.

\section{Conclusions}
\label{sec:conc}

It has been argued that galaxies simulated within a  $\Lambda$CDM universe are not able to 
reproduce the observed diversity of dwarf's RCs, which
may represent a problem for the \lcdm\ cosmological model \citep{Oman15}. We have made a  comparison between circular  velocities of simulated galaxies from the NIHAO project 
and
observed rotational velocities of  galaxies belonging to
the new SPARC dataset 
and to the LITTLE THINGS survey,
spanning a wide range of masses  from  Milky Way to dwarf galaxies,
to check whether simulations that include baryonic processes that modify the distribution of mass in the central regions of DM halos can reproduce the variety of observed RC shapes.

We derived a method to 
measure
the circular rotational velocity
 at the `same' radius for both observed and simulated  galaxies of comparable stellar mass.
We  then compared the circular velocities at the inner 2 kpc and at the outermost radii ($R_{\rm last}$).
 This $V_{\rm 2kpc}-V_{\rm Rlast}$ relation gives an idea of the global shape of the RC.
 The main results of this paper are:

\begin{itemize}

\item
The trend of RC shapes as a function of $V_{\rm Rlast}$ is very similar for simulated NIHAO galaxies and observed  galaxies:
the largest deviation from DM-only predictions occurs at   60$<V_{\rm Rlast}$/\kms$<$100 for both observations and NIHAO simulations, corresponding to  the
$M_{\rm star}\sim$10$^{8.0-9.0}$ M$_\odot$
 mass range,
at which core formation from stellar feedback has been previously shown to be most efficient 
\citep[e.g.,][]{DiCintio14a, Tollet16}.

\item
NIHAO has significantly greater  scatter in  RC shapes compared to previous studies
that do not foresee core creation from stellar feedback, bringing simulation results into better agreement with observations:
the largest scatter in RC shapes for both observations and simulations   is indeed found in the mass range where core formation is most efficient.

\item
Observed galaxies such as
IC~2574 and UGC~5750, with extremely shallow-rising RCs that represented a long-standing issue for \lcdm, are now very well matched by NIHAO simulations, once observational uncertainties are considered and the actual
gradient of the potential is used to compute circular velocities.

\item 
 Some observational outliers are found outside the  $\pm$3$\sigma$ contours covered by NIHAO galaxies in the $V_{\rm 2kpc}-V_{\rm Rlast}$ plane:
the majority of these outlier galaxies are starburst
galaxies or present emission lines that may indicate significant ongoing
star formation activity.
Remarkably, their RC is too steep even when compared to pure N-body simulations that predict a cuspy profile,
indicating a tight link between the  starbursting activity and the central dynamical mass of these galaxies. 

\end{itemize}

Some extra degree of scatter in observed RC shapes  is expected and comes from observational issues: indeed, 
8 out of 11 observed outliers have
poor quality 
 RCs,
which  
   denotes problems  such as disturbed H\,{\sc i} discs containing holes or being warped, significant uncertainties in distances and inclinations, abrupt changes of inclination throughout the disc, insufficient angular resolution over the beam, 
    asymmetries and non-circular or random motions.

%
The two most cored  observed SPARC galaxies
 are outliers when using the spherical circular velocities,
but result very much recovered when using the true circular
velocity criterion, which takes into account the real mass distribution of the
simulated galaxy and is computed from the gravitational potential
at the disc midplane. For galaxies in equilibrium, this is in fact
the same velocity than that measured from the rotation of the H\,{\sc i}
gas, therefore allowing a more consistent comparison with observational
data.
%

The  remaining  
 outlier galaxies all have RCs that rise fast in the inner regions
and then show a sharp change to flat rotation,
contrary to all simulated galaxies  in the $V_{\rm Rlast}\sim$50-100 \kms\ range  
which instead 
have a  smooth RC
and present a central  DM core of various sizes.
We note that
some of these galaxies have even higher inner velocities than those of
dwarf galaxies from simulations where DM halos are barely affected by baryonic processes, and whose RCs therefore reflect their  initial NFW DM distributions.
While 
some objects have  low-quality H\,{\sc i} velocity data --meaning that they are out of equilibrium and their H\,{\sc i} velocities do not trace the underlying potential--
 in the case of  the high quality data, 
such quickly rising and flat RCs
 can be 
 indicating that the inner baryonic mass fraction cannot be neglected, 
  and their halos may have been adiabatically contracted.

In fact, the majority of such observed steeply-rising outlier galaxies are either starbursts 
\citep[see][]{Read16} 
or emission-line galaxies. 
In agreement with previous studies that suggest that central starburst activity is closely linked to 
a high central dynamical mass in dwarf galaxies,
these particular outliers tend to have a lower-than-average effective radius for galaxies of their stellar mass 
\citep[see fig.2 in][]{Lelli16}.
 While the NIHAO galaxies can reproduce well the general trends of the observed SPARC  $M_{\rm star}-r_{\rm eff}$ relation
 \citep{Dutton16},
  they cannot reproduce such compact objects with short effective radii, which  appear to be outliers not only when compared to our simulations, but also when compared to the average population of galaxies of similar stellar mass
 (see compilation of effective radii and $M_{\rm star}$
 data set in Lelli et al.)

 Thus,  our analysis shows that the galaxies with the most compact stellar populations 
 are the ones that have retained a 
cuspy, or possibly contracted, central halo
  profile:
  this  is interesting since  baryon expansion effects are predicted to apply to both stellar and DM \citep{Teyssier13,Dutton16,DiCintio17}. 
 The implication is that this  population of starburst galaxies may have had a mass accretion and star formation history that minimized any baryon-driven expansion  effects. 
Furthermore, it is worth noting that  the exceptionally high rate of star formation that defines starburst galaxies is often triggered by interactions and mergers \citep{Lelli14c}. This is inferred from their RCs, frequently out of equilibrium.
 The NIHAO project, on the contrary, only simulates isolated galaxies, meaning that it is perhaps not unexpected that they do not reproduce any starburst galaxy, and therefore do not match their observed RCs.

Although  NIHAO simulations do not perfectly reproduce observations, their ability to match the general trends in observed RC shapes is encouraging, particularly the match to the features in the velocity region where core formation is predicted to be most efficient. Most observed  dwarf galaxies' RCs in the $V_{\rm Rlast}\sim$60--100 \kms\ range
(corresponding to $M_{\rm star}\sim 10^{8-9}$ M$_\odot$) indicate the presence of a DM core, whilst most galaxies outside this range show a  cuspy DM profile, just as predicted by theoretical models that take into account the effect of gas outflows on DM halos \citep[e.g.,][]{Governato12}.
Observations suggest that the implied core sizes vary greatly in extent even in galaxies of similar total mass. 
Different hydrodynamical simulations
with realistic feedback models
 have shown that this can be explained by the $M_{\rm star}/M_{\rm halo}$-dependent core formation mechanism  \citep{DiCintio14a,chan15,Tollet16}
where at a given $M_{\rm halo}$,   variations  in  $M_{\rm star}$ produce cores of different sizes.
Simulations in \lcdm\  are therefore able to reproduce the expected diversity of RC shapes, without resorting to alternative DM scenarios \citep[e.g.,][]{Kamada16,Creasey17}.
Similar results are also obtained with semi-empirical models just by accounting for  the scatter in the $M_{\rm star}-M_{\rm halo}$  and $M_{\rm star}-r_{\rm eff}$ relations \citep[see][]{brook15a}.


 By contrast, explaining the inability of NFW profiles to match the observed trends 
in RC shapes 
  by invoking observational errors
   would imply that  
i) in the $\sim 50-100$ \kms\ $V_{\rm Rlast}$ range,  80 per cent of observed galaxies have  errors in their RC measurements that preferentially bias them toward showing a core: this means that the error must be `systematic',  favoring the existence of  cores, which is the mean observational trend;
 and ii) such systematic errors must only occur at $V_{\rm max}\approx V_{\rm Rlast}\sim 50-100$ \kms, but not at higher or lower  $V_{\rm max}$.  While both circumstances cannot be ruled out 
\citep[see e.g.][]{Oman17},
  they would point toward some sort of \textit{conspiracy for which observational errors  mimick the presence of a DM core exactly in the range where we  expect DM cores from theoretical models.}
 
 We conclude that  DM halo expansion within a standard \lcdm\ cosmological framework is a viable explanation for the  diversity of observed dwarf galaxy RCs.

\section*{Acknowledgements}
The authors thank the referee Kyle Oman for a critical review of this manuscript.
They also thank Federico Lelli for providing the SPARC data in electronic format and for fruitful discussions. Computational resources  were provided by the High Performance Computing at NYUAD, the  {\sc theo} cluster at MPIA and  the {\sc hydra} clusters  at Rechenzentrum in Garching. 
This work was partially supported by MINECO/FEDER (Spain)  grant AYA2015-63810-P.
 ADC is supported by the Karl Schwarzschild independent fellowship program. 
CBB thanks MINECO/FEDER grant AYA2015-63810-P and the Ramon y Cajal fellowship.
 AD is supported by grants ISF 124/12, BSF 2014-273.

\bibliographystyle{mn2e}
\bibliography{archive}


\label{lastpage}

\end{document}